\documentclass[aps,prd,showpacs,showkeys]{revtex4}
\usepackage{amssymb}
\usepackage{amsfonts}
\usepackage{amsmath}
\usepackage{graphicx}
\usepackage{color}
\usepackage[dvips]{epsfig}
\usepackage[dvips]{graphicx}
\usepackage{float}
\begin{document}

\title{$SQED_{4}$ and $QED_{4}$ on the null-plane}
\author{R. Casana}\email{rodolfo.casana@gmail.com}
\affiliation{$^{1}$Departamento de F\'{\i}sica, Universidade Federal do Maranh\~ao (UFMA) Campus Universit\'ario do Bacanga, CEP 65085-580, S\~ao
Lu\'{\i}s, MA, Brazil.}

\author{B. M. Pimentel} \email{pimentel@ift.unesp.br}
\affiliation{$^{2}$ Instituto de F\'{\i}sica Te\'orica (IFT/UNESP), UNESP - S\~ao Paulo, State University Rua Pamplona 145, CEP 01405-900, S\~ao Paulo,
SP, Brazil.} 

\author{G. E. R. Zambrano} \email{gramos@udenar.edu.co}
\affiliation{$^{3}$Departamento de F\'{\i}sica, Universidad de Nari\~{n}o,
              Ciudad Universitaria Torobajo - Clle 18 Cr 50, San Juan de Pasto - Nari\~{n}o - Colombia.}

\begin{abstract}
We studied the scalar electrodynamics ($SQED_{4}$) and the spinor
electrodynamics ($QED_{4}$) in the null-plane formalism. We followed the
Dirac's technique for constrained systems to perform a detailed analysis of
the constraint structure in both theories. We imposed the appropriated
boundary conditions on the fields to fix the hidden subset first class
constraints which generate improper gauge transformations and obtain an
unique inverse of the second class constraint matrix. Finally, choosing the
null-plane gauge condition, we determined the generalized Dirac brackets of
the independent dynamical variables which via the correspondence principle
give the (anti)-commutators for posterior quantization.
\end{abstract}


 
\maketitle

\section{Introduction}

Half the of last century Dirac \cite{[1]} proposed three different forms of
relativistic dynamics depending on the types of surfaces where the
independent degree of freedom was initiated. The first one, named the
\textit{instant form}, is when a space-like surface is chosen to establish
the fundamental Poisson brackets or commutations relations. It has been used
most frequently so far and is usually called equal-time quantization. The
second form, the \textit{point form}, is to take a branch of hyperbolic
surface $x^{\mu }x_{\mu }=\kappa ^{2}$~with~$x^{0}>0$. And, the third form,
\textit{front form or light front}, is when we choose the surface of a
single light wave to study the field dynamics; it is commonly referred as
the \textit{null-plane formalism} and it took almost 30 years for Dirac's
idea was applied in physical phenomena. An important advantage pointed out
by Dirac is that seven of the ten Poincar\'{e} generators are kinematical on
the null-plane while in the conventional theory constructed on the instant
form only six have this property. Other notable feature of a relativistic
theory on the null-plane is that it gives origin to singular Lagrangians,
e.g. constrained dynamical systems, thus, Dirac's procedure \cite{[2]} can
be employed to analyze the constraint structure of a given theory. In
general, it leads to a reduction in the number of independent field
operators in the respective phase space.

At equal-time, any two different points are space-like separated and
therefore the fields defined at these points are naturally independent
quantities. In a null-plane surface the situation is different because the
micro-causality principle leads to locality requirement in which only the
transversal components are and the longitudinal component becomes non-local
in the theory, although, such situation would not be unexpected \cite{[4]}.
It is possible to verify that the transformation from the usual coordinates
to the null-plane coordinates is not a Lorentz transformation and the
structure of the phase space is different when we compare with the
conventional one. As such, the description of a physical system in the
null-plane formalism could give additional information from those provided
by the conventional formalism \cite{[4]}.   For example, the momentum four-vector
is $(k^+, k^-, k^T)$ where $k^+$ is the null-plane energy while $k^T$ and $k^-$ indicates the transverse and
the longitudinal components of the momentum. Therefore, a massive particle on the mass shell, $k^-=\frac{m^2+(k^T)^2}{2k^+}$,  has positive definite values for $k\pm$ in contrast to $-\infty\leq k^{1,2,3}\leq$ for the usual components. An immediate consequence is that the vacuum on the null-plane
quantized theory may become simpler than the one in the conventional (equal-time) theory
and in many cases the interacting theory vacuum on the null-plane may be the same as the
perturbation theory vacuum. For example, the conservation of the total longitudinal
momentum would not permit the excitations of particle-antiparticle pairs by the null-plane
vacuum (having $k^+ = 0$) \cite{[4a]}.

It was observed that the quantization in the null-plane means to perform the
quantization on the characteristic surfaces of the classical field
equations. Thus, it implies that one has to specify the Cauchy data on both
characteristics, $x^{+}=cte$ and $x^{-}=cte$, and not only on one simple
null-plane \cite{[5]}. In this context, the light cone quantization of free
massless fermions in $(1+1)$--dimensions on both characteristics shows that
the procedure leads to the correct physical descriptions \cite{[6]}.

On the other hand, in \cite{[7]} it was showed an important problem
associated with the quantization on the null-plane:  After establishing the
gauge fixing condition for the first class constraints and the second class
constraints have seen handled through the Dirac's procedure, no more proper
gauge transformations, correspondent to first class constraints, can be performed; however, it still remains in the
analysis a species of improper gauge transformations  associated to the existence of hidden first class constraints and which are related with the zero
mode of the longitudinal derivative $\partial _{-}$ and appears due the lack
of appropriate boundary conditions over the fields \cite{[8]}. This fact
does not allow to define a unique inverse for the second class constraint
matrix that is used to define consistent  Dirac brackets (DB), therefore, the
improper gauge transformations must be fixed by imposing appropriate
boundary conditions.

The present work is addressed to study the constraints structure of the
scalar and the spinor electrodynamics on the null-plane following the
Dirac's formalism for constrained systems. The paper is organized as follow:
In the section \textbf{2} we study the $SQED_{4}$, its constraints structure
is analyzed in detail and the appropriate equations of motion of the
dynamical variables is determined using the extended Hamiltonian. We
classify the set of constraints and find that one of the first class
constraints is a linear combination of scalar and electromagnetic
constraints which is a null vector of the respective constraint matrix. We
invert the first class constraints with the corresponding gauge conditions
and an unique inverse of the matrix of second class constraints is getting
by imposing appropriate boundary conditions on the fields which eliminate
the hidden first class constraints and next we calculate the Dirac's
brackets among the fundamental dynamical variables. In section \textbf{3} we
study the $QED_{4}$, showing that the use of the projection of the fermionic
fields allows to observe the existence of only second class constraints and
the first class constraints in the fermionic sector are associated with the
hidden subset of first class constraints which generate improper gauge
transformations. We also show that the fermionic constraints determine that
the electron field is fully described by only two of the four components. We
use the null-plane gauge to transform the set of first class constraints in
second class and we obtain a graded algebra imposing boundary conditions on
the independent components. In the last section we give our remarks and
conclusions.

\section{Scalar Electrodynamics ($SQED_{4}$): Constraint structure}

The gauge theory we are considering is defined by the following Lagrangian
density in $4$-dimensional space-time%
\begin{equation}
\mathcal{L}=\mathcal{-}\frac{1}{4}F^{\mu \nu }F_{\mu \nu }+g^{\mu \nu
}D_{\mu }\phi \left( D_{\nu }\phi \right) ^{\ast }-m^{2}\phi \phi ^{\ast },
\label{CS.1}
\end{equation}%
here $\phi $ is a one-component complex scalar field, $F_{\mu \nu }\equiv
\partial _{\mu }A_{\nu }-\partial _{\nu }A_{\mu }$ and $D_{\mu }\equiv
\partial _{\mu }+igA_{\mu }$ is the covariant derivative. \ The model is
invariant under the following local $U(1)$ gauge symmetry%
\begin{equation}
\phi \rightarrow e^{i\alpha \left( x\right) }\phi ~\ \ \ ,~\ \ \ \phi ^{\ast
}\rightarrow e^{-i\alpha \left( x\right) }\phi ^{\ast }~\ \ \ ,~\ \ \ A_{\mu
}\rightarrow A_{\mu }-\frac{1}{g}\partial _{\mu }\alpha .
\end{equation}

The field equations are given for%
\begin{equation}
\partial _{\mu }F^{\mu \alpha }+j^{\alpha }=0
\end{equation}%
and
\begin{equation}
\left( D_{\mu }^{\ast }D^{\ast \mu }+m^{2}\right) \phi ^{\ast }=0~\ ~\ \ \ \
,~\ ~\ \ \ \left( D_{\mu }D^{\mu }+m^{2}\right) \phi =0
\end{equation}%
where $j^{\alpha }$ is the current defined by%
\begin{equation}
j^{\mu }\equiv ig\left[ \smallskip \phi \left( \partial ^{\mu }\phi ^{\ast
}-igA^{\mu }\phi ^{\ast }\right) -\phi ^{\ast }\left( \partial ^{\mu }\phi
+igA^{\mu }\phi \right) \right] .  \label{CS.3}
\end{equation}

The canonical conjugate momenta of the fields $A_{\mu },~\phi $ and$~\phi
^{\ast }$ are
\begin{equation}
\pi ^{\mu }\equiv \frac{\partial \mathcal{L}}{\partial \left( \partial
_{+}A_{\mu }\right) }=F^{\mu +},  \label{cs.7a}
\end{equation}%
\begin{equation}
\pi ^{\ast }\equiv \frac{\partial \mathcal{L}}{\partial \left( \partial
_{+}\phi \right) }=\left( D_{-}\phi \right) ^{\ast }~\ ~\ \ \ \ ,~\ ~\ \ \
\pi \equiv \frac{\partial \mathcal{L}}{\partial \left( \partial _{+}\phi
^{\ast }\right) }=D_{-}\phi  \label{cs.7b1}
\end{equation}%
respectively. Then, from (\ref{cs.7a}) and (\ref{cs.7b1}) we get one
dynamical relation%
\begin{equation}
\pi ^{-}=\partial _{+}A_{-}-\partial _{-}A_{+}  \label{cs.8}
\end{equation}%
and five primary constraints, three for the electromagnetic sector
\begin{equation}
\smallskip C\equiv \pi ^{+}\approx 0~\ \ \ \ ,~\ \ \ \ \ \ \ \ \smallskip
\chi ^{k}\equiv \pi ^{k}-\partial _{-}A_{k}+\partial _{k}A_{-}\approx 0
\label{CS.7a}
\end{equation}%
and, two for the scalar sector%
\begin{equation}
\smallskip \Gamma \equiv \pi -D_{-}\phi \approx 0~\ \ \ \ ,~\ \ \ \ \ \ \ \
\smallskip \Gamma ^{\ast }\equiv \pi ^{\ast }-\left( D_{-}\phi \right)
^{\ast }\approx 0.  \label{CS.7b}
\end{equation}

Following the Dirac's procedure \cite{[2]}, we define the canonical
Hamiltonian density which is given by
\begin{equation}
\mathcal{H}_{C}=\frac{1}{2}\left( \pi ^{-}\right) ^{2}+\left( \pi
^{-}\partial _{-}+\pi ^{k}\partial _{k}-j^{+}\right) A_{+}-D_{k}\phi \left(
D^{k}\phi \right) ^{\ast }+m^{2}\phi \phi ^{\ast }+\frac{1}{4}F_{kl}F_{kl}\;,
\label{CS.10}
\end{equation}%
consequently, the canonical Hamiltonian is $H_{C}=\displaystyle\int \!d^{3}y~%
\mathcal{H}_{C}$, with $\displaystyle\int \!d^{3}y=\displaystyle\int
\!dy^{1}dy^{2}dy^{-}$.

We also define the primary Hamiltonian $H_{P}$ adding to the canonical
Hamiltonian the primary constraints with their respective Lagrange
multipliers
\begin{equation}
H_{P}=H_{C}+\int \!\!d^{3}y\left( \mathrm{w_1}C+\mathrm{u}_{k}\chi ^{k}+%
\mathrm{v}^{\ast }\Gamma +\Gamma ^{\ast }\mathrm{v}\right) ,  \label{CS.11}
\end{equation}%
where $\mathrm{w}_1$, $\mathrm{u}_{k}$ are the multipliers related to the
electromagnetic constraints and, $\mathrm{v}$ and $\mathrm{v}^{\ast }$ are
the respective multipliers for the scalar constraints.

The fundamental Poisson brackets (PB) between the fields are
\begin{equation}
\left\{ A_{\mu }\left( x\right) ,\pi ^{\mu }\left( y\right) \right\} =\delta
_{\mu }^{\nu }\delta ^{3}\left( x-y\right) ,
\end{equation}%
\begin{equation}
\left\{ \phi \left( x\right) ,\pi ^{\ast }\left( y\right) \right\} =\delta
^{3}\left( x-y\right) ~\ \ \ \ ,~\ \ \ \left\{ \phi ^{\ast }\left( x\right)
,\pi \left( y\right) \right\} =\delta ^{3}\left( x-y\right) ~\ .
\end{equation}

The Dirac's procedure tell us that the primary constraints must be preserved
in time (consistence condition) under time evolution generated by the
primary Hamiltonian by requiring that they have a weakly vanishing PB with $%
H_{P}$. Thus, such requirement on the scalar constraints yields
\begin{eqnarray}
\dot{\Gamma} &=&-ig\left[ \phi \pi ^{-}+2D_{-}\left( A_{+}\phi \right) %
\right] -D^{k}D_{k}\phi -m^{2}\phi -2D_{-}\mathrm{v}\approx 0,  \notag \\%
[-0.2cm]
&&  \label{CS.13} \\[-0.15cm]
\dot{\Gamma}^{\ast } &=&ig\left[ \phi ^{\ast }\pi ^{-}+2D_{-}^{\ast }\left(
A_{+}\phi ^{\ast }\right) \right] -\left( D_{k}D^{k}\phi \right) ^{\ast
}-m^{2}\phi ^{\ast }-2\left( D_{-}\mathrm{v}\right) ^{\ast }\approx 0.
\notag
\end{eqnarray}%
These relations determine the multipliers $\mathrm{v}$ and $\mathrm{v}$,
respectively, and there are not more constraints associate with the scalar
sector.

In the electromagnetic sector, the consistence condition of $\chi ^{k}$
yields
\begin{equation}
\dot{\chi}^{k}=\partial _{k}\pi ^{-}+j^{k}+\partial _{j}F_{jk}-2\partial _{-}%
\mathrm{u}_{k}\approx 0,  \label{CS.16}
\end{equation}%
an equation for its associated multiplier $\mathrm{u}_{k}$.

Finally, the consistence condition of $\pi ^{+}$ gives a secondary
constraint $G$
\begin{equation}
\dot{C}=G=\partial _{-}\pi ^{-}+\partial _{k}\pi ^{k}+j^{+}\approx 0,
\label{CS.14}
\end{equation}%
which is simply the Gauss's law. It is easy to verify that there are not
further constraints generate from the consistence condition of the Gauss's
law because it is automatically conserved
\begin{equation}
\dot{G}\equiv ig\left[ \phi ^{\ast }\dot{\Gamma}-\phi \dot{\Gamma}^{\ast }%
\right] \approx 0.  \label{CS.15}
\end{equation}

\smallskip

Then, there are not more constraints in the theory and the full set of
constraints given by (\ref{CS.7a}), (\ref{CS.7b}) and (\ref{CS.14}).

\subsection{Constraint classification}

The constraint $\pi ^{+}$ has vanishing PB with all the other constraints
therefore it is a first class constraint. The remaining set $\Phi
^{a}=\left\{ \Gamma ,~\Gamma ^{\ast },~G,~\chi ^{k}\right\} $ is apparently
a second class set, however, the determinant of its constraint matrix $%
\left\{ \Phi ^{a}\left( x\right) ,\Phi ^{b}\left( y\right) \right\} $ is
zero (see Appendix C) because the matrix has a zero mode whose eigenvector
gives a linear combination of constraints which is one more first class
constraint. Alternately, there is an additional argument, the constraint $G$%
, the Gauss's law, is the second first class constraint for zero coupling
constant: the free Maxwell's field theory. On the other hand, if $G$ belongs
to a minimal set of second class constraints, the limit for zero coupling
constant would no longer be possible because the DB would become undefined.
Remembering that the DB is defined with respect to a non-singular matrix,
however, it could become singular when we go back to the free
theory. In conclusion, there is a linear combination which is independent of
$\pi ^{+}$ and it is a first class constraint. Such that linear combination
is
\begin{equation}
\Sigma \equiv G-ig\left( \phi ^{\ast }\Gamma -\phi \Gamma ^{\ast }\right) .
\label{CS.17}
\end{equation}

Thus, we have the following set of second class constraints
\begin{eqnarray}
\Gamma &\equiv &\pi -D_{-}\phi \approx 0\qquad ,\qquad \Gamma ^{\ast }\equiv
\pi ^{\ast }-\left( D_{-}\phi \right) ^{\ast }\approx 0\quad ,
\label{CS.19a} \\
\chi ^{k} &\equiv &\pi ^{k}-\partial _{-}A_{k}+\partial _{k}A_{-}\approx 0
\notag
\end{eqnarray}%
and the set of first class constraints
\begin{equation}
C=\pi ^{+}\approx 0\qquad ,\qquad \Sigma =G-ig\left( \phi ^{\ast }\Gamma
-\phi \Gamma ^{\ast }\right) \approx 0.  \label{CS.20}
\end{equation}

We can be sure that is the maximal number of first class constraints since
the quest for time independence of $\left( \Gamma ,~\Gamma ^{\ast },~\chi
^{k}\right) $ leads to equations that determine their respective Lagrange
multipliers. The second first class constraint in (\ref{CS.17}), must be
contrasted with the \emph{instant form} analysis \cite{[9]} where the second
first class constraint is not a linear combination of electromagnetic and
scalar constraints because in this formalism the scalar sector does not have
any constraint.

\subsection{Equations of motion and gauge fixing conditions}

At this point we need to check that we have the correct (Euler-Lagrange)
equations of motion. Thus, the time derivative of the fields is calculated
by computing their PB with the so called extended Hamiltonian ($H_{E}$) what
now it generates the time translations or temporal evolutions. The $H_{E}$
is obtained by adding to the primary Hamiltonian $H_{P}$ all the first class
constraints, thus, we get
\begin{eqnarray}
H_{E} &=&\int \!\!d^{3}y\left[ \frac{1}{2}\left( \pi ^{-}\right) ^{2}+\left(
\pi ^{-}\partial _{-}+\pi ^{k}\partial _{k}-j^{+}\right) A_{+}-D_{k}\phi
\left( D^{k}\phi \right) ^{\ast }+m^{2}\phi \phi ^{\ast }\right.  \notag \\
&&  \label{CS.22} \\[-0.2cm]
&&\left. +\frac{1}{4}F_{kl}F_{kl}\right] +\int \!\!d^{3}y\left[ \frac{{}}{{}}%
\mathrm{w}_{1}C+\mathrm{u}_{k}\chi ^{k}+\mathrm{v}^{\ast }\Gamma +\Gamma
^{\ast }\mathrm{v+w}_{2}\Sigma \right] .  \notag
\end{eqnarray}

Thus we have that the time evolution of the dynamical variables of the
electromagnetic field are%
\begin{equation*}
\begin{array}{lllll}
\dot{A}_{+}=\mathrm{w}_{1} & \quad & , & \quad & \dot{\pi}^{+}=G\approx 0 \\
&  &  &  &  \\
\dot{A}_{-}=\pi ^{-}+\partial _{-}A_{+}-\partial _{-}\mathrm{w}_{2} &  & , &
& \dot{\pi}^{-}=2g^{2}\phi \phi ^{\ast }A_{+}+\partial _{k}\mathrm{u}_{k}+ig%
\mathrm{v}^{\ast }\phi -ig\phi ^{\ast }\mathrm{v} \\
&  &  &  &  \\
\dot{A}_{k}=\partial _{k}A_{+}+\mathrm{u}_{k}-\partial _{k}\mathrm{w}_{2} &
& , &  & \dot{\pi}^{k}=ig\phi \left( D^{k}\phi \right) ^{\ast }-ig\phi
^{\ast }D^{k}\phi +\partial _{j}F_{jk}-\partial _{-}\mathrm{u}_{k}%
\end{array}%
\end{equation*}%
and for the scalar fields are given by%
\begin{eqnarray*}
\dot{\phi} &=&\mathrm{v}+ig\mathrm{w}_{2}\phi \\
\dot{\phi}^{\ast } &=&\mathrm{v}^{\ast }-ig\mathrm{w}_{2}\phi ^{\ast } \\
\dot{\pi}^{\ast } &=&-\left( D_{-}D_{+}\phi \right) ^{\ast }+igA_{+}\left(
D_{-}\phi \right) ^{\ast }-\left( D_{k}D^{k}\phi \right) ^{\ast }-m^{2}\phi
^{\ast }-ig\phi ^{\ast }D_{-}^{\ast }\mathrm{w}_{2} \\
&&-2ig\mathrm{w}_{2}\left( D_{-}\phi \right) ^{\ast } \\
\dot{\pi} &=&-D_{-}D_{+}\phi -igA_{+}D_{-}\phi -D^{k}D_{k}\phi -m^{2}\phi
+ig\phi D_{-}\mathrm{w}_{2}+2ig\mathrm{w}_{2}D_{-}\phi .
\end{eqnarray*}%
From which, it is easy to obtain
\begin{eqnarray}
D_{\mu }D^{\mu }\phi +m^{2}\phi &\approx &ig\phi D_{-}\mathrm{w}_{2}+2ig%
\mathrm{w}_{2}D_{-}\phi \;,  \notag \\[0.25cm]
\left( D_{\mu }D^{\mu }\phi \right) ^{\ast }+m^{2}\phi ^{\ast } &\approx
&-ig\phi ^{\ast }D_{-}^{\ast }\mathrm{w}_{2}-2ig\mathrm{w}_{2}\left(
D_{-}\phi \right) ^{\ast }\;,  \label{CS.23} \\[0.25cm]
\partial _{\mu }F^{\mu \nu }+j^{\nu } &\approx &0\;,  \notag
\end{eqnarray}%
thus, the equation of motions are consistent with its Lagrangian form if we
choose $\mathrm{w}_{2}=0$.

The Dirac's algorithm requires as many gauge conditions as first class
constraints there are. However, such gauge fixing conditions must be
compatible with the Euler-Lagrange equations and therefore they must fix the
Lagrangian multiplier $\mathrm{w}_{2}$ to zero and together with the first
class constraints must be a second class set. One set of gauge conditions
satisfying such requirements is
\begin{equation}
A_{-}\approx 0\qquad ,\qquad \pi ^{-}+\partial _{-}A_{+}\approx 0
\label{CS.21}
\end{equation}%
which are standard in the pure gauge theory, the so-called null-plane gauge
\cite{[7],[10]}.

\subsection{Dirac's brackets (DB)}

We follow the iterative method to calculate the Dirac's brackets, thus, we
first consider the set of the first class constraints (\ref{CS.20}) and
their gauge fixing conditions (\ref{CS.21})
\begin{equation}
\begin{array}{lllll}
\Phi _{1}\equiv \pi ^{+} & \quad & , & \quad & \Phi _{3}\equiv A_{-} \\
&  &  &  &  \\
\Phi _{2}\equiv \Sigma =G-ig\left( \phi ^{\ast }\Gamma -\phi \Gamma ^{\ast
}\right) &  & , &  & \Phi _{4}\equiv \pi ^{-}+\partial _{-}A_{+},%
\end{array}
\label{IC.1}
\end{equation}%
such that the set of constraints \eqref{IC.1} is second class and whose constraint matrix $C_{ij}\left( x,y\right) \equiv \left\{ \Phi
_{i}\left( x\right) ,\Phi _{j}\left( y\right) \right\} $, with components: 
\begin{equation}
C_{ij}\left( x,y\right) =\left(
\begin{array}{cccc}
0 & 0 & 0 & \partial _{-}^{x} \\[0.2cm]
0 & 0 & -\partial _{-}^{x} & 0 \\[0.2cm]
0 & -\partial _{-}^{x} & 0 & 1 \\[0.2cm]
\partial _{-}^{x} & 0 & -1 & 0%
\end{array}%
\right) \delta ^{3}\left( x-y\right) .  \label{IC.2}
\end{equation}
must be regular. However, in order to solve the inverse of the constraint matrix  \eqref{IC.2} we require a suitable
inverse for the longitudinal derivative $\partial _{-}$. In general, the operator $\partial _{-}$ has the following inverse:
\begin{equation}
\left( \partial _{-}\right) ^{-1}f(x^{-})=\frac{1}{2}\int
\!\!dy^{-}~\epsilon \left( x^{-}-y^{-}\right) f\left( y^{-},x^{+},x^{\bot
}\right) +F\left( x^{+},x^{\bot }\right) ,  \label{IC.3}
\end{equation}%
where the function $\epsilon \left( x\right) $ is
\begin{equation}
\epsilon \left( x\right) =\left\{
\begin{array}{rrr}
1 & , & x>0 \\
0 & , & x=0 \\
-1 & , & x<0%
\end{array}%
\right.
\end{equation}%
and $F\left( x^{+},x^{\bot }\right) $ is a  $x^{-}$ arbitrary independent function. The
presence of $F$  implies that constraint matrix \eqref{IC.2} does not have a unique
inverse, nevertheless, Dirac proved that the matrix formed by a complete set of second
class constraints should be unique, therefore, it is said that the set of second class
constraints in (\ref{IC.1}) is not purely second class. 

Steinhardt \cite{[7]} proved that the inverse matrix
of (\ref{IC.2}) is not unique because among the second class constraints
there are a hidden subset first class constraints \cite{[8]}. This subset of  constraints can be evidenced  by observing that the most general solution for
 (\ref{CS.13}) and (\ref{CS.16}) is:
\begin{eqnarray}
\mathrm{v}\left( x\right) &=&\hat{\mathrm{v}}\left( x\right) +s\left(
x^{+},x^{\bot }\right) ,  \notag \\[0.1in]
\mathrm{v}^{\ast }\left( x\right) &=&\hat{\mathrm{v}}^{\ast }\left( x\right)
+s^{\ast }\left( x^{+},x^{\bot }\right) ,  \label{V} \\[0.1in]
\mathrm{u}_{k}\left( x\right) &=&\hat{\mathrm{u}}_{k}\left( x\right)
+s_{k}\left( x^{+},x^{\bot }\right) ,  \notag
\end{eqnarray}%
where $s\left( x^{+},x^{\bot }\right) $, $s^{\ast }\left( x^{+},x^{\bot
}\right) $ and $s_{k}\left( x^{+},x^{\bot }\right) $ are arbitrary functions
of $\left( x^{+},x^{\bot }\right) $, and $\hat{\mathrm{v}}\left( x\right) $,
$\hat{\mathrm{v}}^{\ast }\left( x\right) $ and $\hat{\mathrm{u}}_{k}\left(
x\right) $ represent the "unambiguous" solutions.  Now, if we insert the $\mathrm{v}$, $\mathrm{v}^{\ast }$ and $\mathrm{u}_{k}$ of \eqref{V} into the extended Hamiltonian  \eqref{CS.22}, then
\begin{eqnarray}
H_{E}^{\prime } &=&\int \!\!d^{3}y\left[ \frac{1}{2}\left( \pi ^{-}\right)
^{2}+\left( \pi ^{-}\partial _{-}+\pi ^{k}\partial _{k}-j^{+}\right)
A_{+}-D_{k}\phi \left( D^{k}\phi \right) ^{\ast }+m^{2}\phi \phi ^{\ast
}\right.  \notag \\
&&\left. +\frac{1}{4}F_{kl}F_{kl}\right] +\int \!\!d^{3}y\left[ \frac{{}}{{}}%
\mathrm{w}_{1}C+\hat{\mathrm{u}}_{k}\chi ^{k}+\hat{\mathrm{v}}^{\ast }\Gamma
+\Gamma ^{\ast }\hat{\mathrm{v}}\mathrm{+w}_{2}\Sigma \right]  \label{IC.4}
\\
&&+\int \!\!d^{3}y\left[ \frac{{}}{{}}s_{k}\chi ^{k}+s^{\ast }\Gamma +\Gamma
^{\ast }s\right] .  \notag
\end{eqnarray}%
Thus, despite that the set of constraints \eqref{IC.1} seem to be second class, the multipliers $\mathrm{v}$, $\mathrm{v}^{\ast }$ and $\mathrm{u}_{k}$ are not completely fixed
implying that the Hamiltonian still contains the arbitrary functions  $s$, $s^{\ast }$ and $s_{k}$.

Steinhardt  shown that this hidden subset of first class
constraints is associated with improper gauge transformations \cite{[7]}. An improper gauge transformation can not identifies  with generators of gauge transformations \cite{[8]}, as occurs with the first class constraints  \eqref{CS.20}. Since this kind of constraints are tied to boundary conditions, they can map a given physical solution to another one with different boundary conditions, which is no equivalent to the former \cite{[8]} \cite{[9]}. Therefore, it is not
possible to eliminate the improper gauge transformations by means of gauge
conditions since such procedure would exclude configurations physically
allowed to the system.  This hidden constraints can be eliminated by fixing
appropriated boundary conditions on the fields in order to the total
Hamiltonian be a true generator of time evolution of the physical system.

Thus, in order to evaluate explicitly  the inverse of the matrix of second class
constraints \eqref{IC.2} and ensure its uniqueness we must determine $F\left( x^{+},x^{\bot }\right) $. This function can be evaluated if we impose the appropriate boundary conditions on the fields $\left( \phi ,~\phi ^{\ast },~A_{k}\right) $ given in the reference \cite{[5]}. Under such boundary conditions, the inverse of the
operator $\partial _{-}$ is defined on all integrable functions $f(x^{-})$
which are less singular than $\displaystyle\frac{1}{x^{-}}$ and vanish
faster than $\displaystyle\frac{1}{x^{-}}$ for large $x^{-}$, namely:
\begin{equation}
\left( \partial _{-}\right) ^{-1}f(x^{-})=\frac{1}{2}\int \!\!dy^{-}\epsilon
\left( x^{-}-y^{-}\right) f(y^{-}).  \label{IC.5}
\end{equation}%
With this, we get a unique inverse of \eqref{IC.2} which is given by
\begin{equation}
C_{ij}^{-1}\left( x,y\right) \equiv \frac{1}{2}\left(
\begin{array}{cccc}
0 & -\left\vert x^{-}-y^{-}\right\vert & 0 & \epsilon \left(
x^{-}-y^{-}\right) \\[0.2cm]
\left\vert x^{-}-y^{-}\right\vert & 0 & -\epsilon \left( x^{-}-y^{-}\right)
& 0 \\[0.2cm]
0 & -\epsilon \left( x^{-}-y^{-}\right) & 0 & 0 \\[0.2cm]
\epsilon \left( x^{-}-y^{-}\right) & 0 & 0 & 0%
\end{array}%
\right) \delta ^{2}\left( x^{\bot }-y^{\bot }\right) .  \label{IC.6}
\end{equation}%
Alternatively, we can determine the inverse by insisting that the DB satisfy
Jacobi identities \cite{[11]}, given the same result.

Using the inverse defined by equation (\ref{IC.6}), the first set of DB, $%
\left\{ \mathbf{\cdot },\mathbf{\cdot }\right\} _{D1}$, for given two
dynamical variables $\mathbf{A}\left( x\right) $ and $\mathbf{B}\left(
y\right) $ are calculate by
\begin{eqnarray}
\left\{ \mathbf{A}\left( x\right) ,\mathbf{B}\left( y\right) \right\} _{D1}
&=&\left\{ \mathbf{A}\left( x\right) ,\mathbf{B}\left( y\right) \right\}
-\int \!\!d^{3}ud^{3}v\left\{ \mathbf{A}\left( x\right) ,\Phi _{i}\left(
u\right) \right\}  \label{IC.7} \\
&& \qquad \qquad \qquad \qquad C_{ij}^{-1}\left( u,v\right) \left\{ \Phi
_{j}\left( v\right) ,\mathbf{B}\left( y\right) \right\} .  \notag
\end{eqnarray}%
Thus, the no null DB$_{1}$ are%
\begin{eqnarray}
\left\{ \phi \left( x\right) ,A_{+}\left( y\right) \right\} _{D1} &=&\frac{ig%
}{2}\phi \left( x\right) \left\vert x^{-}-y^{-}\right\vert \delta ^{2}\left(
x^{\bot }-y^{\bot }\right) ,  \notag \\[0.25cm]
\left\{ \phi ^{\ast }\left( x\right) ,A_{+}\left( y\right) \right\} _{D1}
&=&-\frac{ig}{2}\phi ^{\ast }\left( x\right) \left\vert
x^{-}-y^{-}\right\vert \delta ^{2}\left( x^{\bot }-y^{\bot }\right) ,
\label{IC.9} \\[0.25cm]
\left\{ A_{k}\left( x\right) ,A_{+}\left( y\right) \right\} _{D1} &=&-\frac{1%
}{2}\left\vert x^{-}-y^{-}\right\vert \partial _{k}^{x}\delta ^{2}\left(
x^{\bot }-y^{\bot }\right) .  \notag
\end{eqnarray}

Now, following with the  iterative procedure to calculate DB \cite{[9]}, 
 we consider the subset of the remaining second class constraints that
under the brackets DB$_{1}$ are given as 
\begin{equation}
\begin{array}{lllll}
\Psi _{1}\equiv \Gamma =\pi -\partial _{-}\phi &  & , &  & \Psi _{2}\equiv
\Gamma ^{\ast }=\pi ^{\ast }-\partial _{-}\phi ^{\ast } \\
&  &  &  &  \\
\Psi _{3}\equiv \chi ^{1} &  & , &  & \Psi _{4}\equiv \chi ^{2}%
\end{array}%
,  \label{DB.1}
\end{equation}%
where $\chi ^{k}=\pi ^{k}-\partial _{-}A_{k}$. The constraint matrix from this set is defines as
\begin{equation}
D_{ij}\left( x,y\right) \equiv \left\{ \Psi _{i}\left( x\right) ,\Psi
_{j}\left( y\right) \right\} _{D1}.  \label{DB.2}
\end{equation}%
Using the boundary conditions on the fields accepted to calculate \eqref{IC.6}, 
we compute the inverse $D^{-1}$ and thus we define the second set of DB, $\left\{ \cdot ,\cdot \right\} _{D2}$,
\begin{eqnarray}
\left\{ \mathbf{A}\left( x\right) ,\mathbf{B}\left( y\right) \right\} _{D2}
&=&\left\{ \mathbf{A}\left( x\right) ,\mathbf{B}\left( y\right) \right\}
_{D1}-\int \!\!d^{3}ud^{3}v\left\{ \mathbf{A}\left( x\right) ,\Psi
_{i}\left( u\right) \right\} _{D1}  \label{DB.3} \\
&&\qquad \qquad \qquad \qquad D_{ij}^{-1}\left( u,v\right) \left\{ \Psi
_{j}\left( v\right) ,\mathbf{B}\left( y\right) \right\} _{D1}.  \notag
\end{eqnarray}%
Then, we obtain the final DB among the fundamental dynamical variables of
the theory
\begin{eqnarray}
\left\{ A_{k}\left( x\right) ,A_{l}\left( y\right) \right\} _{D2} &=&-\frac{1%
}{4}\delta _{l}^{k}\epsilon \left( x^{-}-y^{-}\right) \delta ^{2}\left(
x^{\bot }-y^{\bot }\right) ,  \notag \\
&&  \label{DB.4} \\
\left\{ \phi \left( x\right) ,\phi ^{\ast }\left( y\right) \right\} _{D2}
&=&-\frac{1}{4}\epsilon \left( x^{-}-y^{-}\right) \delta ^{2}\left( x^{\bot
}-y^{\bot }\right) ,  \notag
\end{eqnarray}%
\begin{eqnarray}
\left\{ \phi \left( x\right) ,A_{+}\left( y\right) \right\} _{D2} &=&\frac{i%
}{2}g\phi \left( x\right) \left\vert x^{-}-y^{-}\right\vert \delta
^{2}\left( x^{\bot }-y^{\bot }\right) -\frac{i}{8}g\delta ^{2}\left( x^{\bot
}-y^{\bot }\right)  \notag \\[0.3cm]
&&\int \!\!dv^{-}~\epsilon \left( x^{-}-v^{-}\right) \phi \left( x^{\bot
},v^{-}\right) \epsilon \left( v^{-}-y^{-}\right) ,  \label{DB.4c}
\end{eqnarray}%
\begin{eqnarray}
\left\{ \phi ^{\ast }\left( x\right) ,A_{+}\left( y\right) \right\} _{D2}
&=&-\frac{i}{2}g\phi ^{\ast }\left( x\right) \left\vert
x^{-}-y^{-}\right\vert \delta ^{2}\left( x^{\bot }-y^{\bot }\right) +\frac{i%
}{8}g\delta ^{2}\left( x^{\bot }-y^{\bot }\right)  \notag \\[0.3cm]
&&\int \!\!dv^{-}~\epsilon \left( x^{-}-v^{-}\right) \phi ^{\ast }\left(
x^{\bot },v^{-}\right) \epsilon \left( v^{-}-y^{-}\right) .  \label{DB.4b}
\end{eqnarray}%
From the correspondence principle we obtain the following commutators among
the fields
\begin{eqnarray}
\left[ \smallskip A_{k}\left( x\right) ,A_{l}\left( y\right) \right] &=&-%
\frac{i}{4}\delta _{l}^{k}\epsilon \left( x^{-}-y^{-}\right) \delta
^{2}\left( x^{\bot }-y^{\bot }\right) ,  \notag \\
&&  \label{A} \\
\left[ \smallskip \phi \left( x\right) ,\phi ^{\ast }\left( y\right) \right]
&=&-\frac{i}{4}\epsilon \left( x^{-}-y^{-}\right) \delta ^{2}\left( x^{\bot
}-y^{\bot }\right) ,  \notag
\end{eqnarray}%
\begin{eqnarray}
\left[ \smallskip \phi \left( x\right) ,A_{+}\left( y\right) \right] &=&-%
\frac{1}{2}g\phi \left( x\right) \left\vert x^{-}-y^{-}\right\vert \delta
^{2}\left( x^{\bot }-y^{\bot }\right) +\frac{1}{8}g\delta ^{2}\left( x^{\bot
}-y^{\bot }\right)  \notag \\[0.12in]
&&\int \!\!dv^{-}~\epsilon \left( x^{-}-v^{-}\right) \phi \left( x^{\bot
},v^{-}\right) \epsilon \left( v^{-}-y^{-}\right) ,  \label{B}
\end{eqnarray}%
\begin{eqnarray}
\left[ \smallskip \phi ^{\ast }\left( x\right) ,A_{+}\left( y\right) \right]
&=&\frac{1}{2}g\phi ^{\ast }\left( x\right) \left\vert
x^{-}-y^{-}\right\vert \delta ^{2}\left( x^{\bot }-y^{\bot }\right) -\frac{1%
}{8}g\delta ^{2}\left( x^{\bot }-y^{\bot }\right)  \notag \\[0.12in]
&&\int \!\!dv^{-}~\epsilon \left( x^{-}-v^{-}\right) \phi ^{\ast }\left(
x^{\bot },v^{-}\right) \epsilon \left( v^{-}-y^{-}\right) .  \label{C}
\end{eqnarray}

The first two relations were specified by Neville and Rohrlich \cite{[12]}.
They derived this expressions starting from the free field operators
commutation relations for unequal times $x^{+}$, $y^{+}$ and then, they
calculated the commutators on the null-plane at equal time, \emph{i. e.}, $%
x^{+}=y^{+}$. The commutation relations involving the field operator $\hat{A}%
_{+}(x)$ were not obtained in \cite{[12]} but it was affirmed that they must
be derived solving a quantum constraint. However, we get to show that it is
possible to obtain these last commutation relations at classical level
following a carefully Dirac's analysis of the constraint structure of the
model.

\section{Spinor Electrodynamics ($QED_{4}$): Constraint structure}

The Lagrangian density of the spinor Electrodynamics written in terms of the
light cone projections\footnote{%
See the definitions in the Appendix \textbf{A}.} of the fermionic fields is
\begin{eqnarray}
\mathcal{L} &=&\bar{\psi}_{+}\left( \frac{i}{2}\gamma ^{+}%
\overleftrightarrow{\partial }_{\!\!+}-gA_{+}\gamma ^{+}\right) \psi _{+}+%
\bar{\psi}_{-}\left( \frac{i}{2}\gamma ^{-}\overleftrightarrow{\partial }%
_{\!\!-}-gA_{-}\gamma ^{-}\right) \psi _{-}  \notag \\[-0.15cm]
&&+\bar{\psi}_{+}\left( \frac{i}{2}\overleftrightarrow{\partial \!\!\!\slash}%
-{g}{A\!\!\!\slash}-m\right) \psi _{-}+\bar{\psi}_{-}\left( \frac{i}{2}%
\overleftrightarrow{\partial \!\!\!\slash}-g{A\!\!\!\slash}-m\right) \psi
_{+}  \label{d64} \\[-0.2cm]
&&\!\!\!-\frac{1}{2}\left( F_{12}\right) ^{2}+\frac{1}{2}\left(
F_{+-}\right) ^{2}+F_{+k}F_{-k}\ ,  \notag
\end{eqnarray}%
where we have defined $\gamma ^{k}A_{k}={A\!\!\!\slash}$ for $k=1,2$. The
corresponding field equations are:
\begin{eqnarray}
\partial _{\nu }F^{\nu \mu }-g\bar{\psi}\gamma ^{\mu }\psi &=&0  \notag \\%
[0.2cm]
\left( i\partial _{+}-gA_{+}\right) \gamma ^{+}\psi _{+}+\left[ i\partial
\!\!\!\slash-g{A\!\!\!\slash}-m\right] \psi _{-} &=&0  \notag \\[0.2cm]
\left( i\partial _{-}-gA_{-}\right) \gamma ^{-}\psi _{-}+\left[ i\partial
\!\!\!\slash-g{A\!\!\!\slash}-m\right] \psi _{+} &=&0  \label{d67} \\[0.2cm]
\left( i\partial _{+}+gA_{+}\right) \bar{\psi}_{+}\gamma ^{+}+\bar{\psi}_{-}%
\left[ i\overleftarrow{\partial \!\!\!\slash}+g{A\!\!\!\slash}+m\right] &=&0
\notag \\[0.2cm]
\left( i\partial _{-}+gA_{-}\right) \bar{\psi}_{-}\gamma ^{-}+\bar{\psi}_{+}%
\left[ i\overleftarrow{\partial \!\!\!\slash}+g{A\!\!\!\slash}+m\right] &=&0.
\notag
\end{eqnarray}%
The canonical momenta for the fields are
\begin{equation}
\pi ^{\mu }=F^{\mu +}=\partial ^{\mu }A^{+}-\partial ^{+}A^{\mu }\;
\label{d7}
\end{equation}%
and
\begin{equation}
\begin{array}{lllll}
\bar{\phi}_{+a}=\displaystyle-\frac{i}{2}\bar{\psi}_{+b}\left( \gamma
^{+}\right) _{ba} &  & , &  & \phi _{+a}=\displaystyle-\frac{i}{2}\left(
\gamma ^{+}\right) _{ab}\psi _{+b}\;, \\
&  &  &  &  \\[-0.2cm]
\bar{\phi}_{-a}=0 &  & , &  & \phi _{-a}=0\;,%
\end{array}
\label{d8}
\end{equation}%
where $a,b=1,2,3,4$.

From the canonical momenta equations, we observe that the only one equation
in (\ref{d7}) is dynamical
\begin{equation}
\pi ^{-}=F^{-+}=\partial _{+}A_{-}-\partial _{-}A_{+},
\end{equation}%
and all the other equations coming from (\ref{d7}) and (\ref{d8}) give a set
of primary constrains composite by three bosonic constraints
\begin{equation}
C\equiv \pi ^{+}\approx 0\quad ~\ ,~\ \ \ C^{k}\equiv \pi ^{k}+\partial
_{k}A_{-}-\partial _{-}A_{k}\approx 0~\ \ ,\ \ ~k=1,2  \label{d68-1}
\end{equation}%
and four fermionic constraints
\begin{equation}
\begin{array}{lllll}
\Gamma _{+a}\equiv \phi _{+a}+\frac{i}{2}\left( \gamma ^{+}\right) _{ab}\psi
_{+b}\approx 0 &  & , &  & \bar{\Gamma}_{+a}\equiv \bar{\phi}_{+a}+\frac{i}{2%
}\bar{\psi}_{+b}\left( \gamma ^{+}\right) _{ba}\approx 0 \\
&  &  &  &  \\
\Gamma _{-a}\equiv \phi _{-a}\approx 0 &  & , &  & \bar{\Gamma}_{-a}\equiv
\bar{\phi}_{-a}\approx 0\ .%
\end{array}
\label{d68-2a}
\end{equation}

The canonical Hamiltonian density is \cite{[13]}
\begin{eqnarray}
\mathcal{H}_{c} &=&\frac{1}{2}\left( \pi ^{-}\right) ^{2}+\left[ \pi
^{-}\partial _{-}+\pi ^{k}\partial _{k}+g\bar{\psi}_{+}\gamma ^{+}\psi _{+}%
\right] A_{+}+\frac{1}{2}\left( F_{12}\right) ^{2}  \label{d69-1} \\[0.1in]
&&-\,\bar{\psi}_{-}\left[ \frac{i}{2}\gamma ^{-}\overleftrightarrow{\partial
}_{\!\!-}-gA_{-}\gamma ^{-}\right] \psi _{-}-\bar{\psi}_{+}\left[ \frac{i}{2}%
\overleftrightarrow{\partial \!\!\!\slash}-g{A\!\!\!\slash}-m\right] \psi
_{-}  \notag \\
&&-\bar{\psi}_{-}\left[ \frac{i}{2}\overleftrightarrow{\partial \!\!\!\slash}%
-g{A\!\!\!\slash}-m\right] \psi _{+},  \notag
\end{eqnarray}%
and the primary Hamiltonian takes the form
\begin{equation}
H_{P}=H_{c}+\int \!\!dy^{3}\left[ uC+u_{k}C^{k}+\bar{\Gamma}_{+a}v_{1a}+\bar{%
\Gamma}_{-a}v_{2a}-\bar{v}_{1a}\Gamma _{+a}-\bar{v}_{2a}\Gamma _{-a}\right] ,
\label{d70}
\end{equation}%
where $u$ and $u_{k}$ are bosonic Lagrange multipliers and, $v_{1},v_{2}$, $%
\bar{v}_{1}$ and $\bar{v}_{2}$ are fermionic multipliers.

The fundamental Poisson brackets are
\begin{eqnarray}
&\left\{ A_{\mu }\left( x\right) ,\pi ^{\nu }\left( y\right) \right\}
=\delta _{\mu }^{\nu }\delta ^{3}\left( x-y\right) ,&  \label{d71} \\[0.3cm]
&\left\{ \psi _{+a}\left( x\right) ,\bar{\phi}_{+b}\left( y\right) \right\}
=-\delta _{ab}\delta ^{3}\left( x-y\right) \quad ,\quad \left\{ \bar{\psi}%
_{+a}\left( x\right) ,\phi _{+b}\left( y\right) \right\} =-\delta
_{ab}\delta ^{3}\left( x-y\right) ,&  \notag \\[0.3cm]
&\left\{ \psi _{-a}\left( x\right) ,\bar{\phi}_{-b}\left( y\right) \right\}
=-\delta _{ab}\delta ^{3}\left( x-y\right) \quad ,\quad \left\{ \bar{\psi}%
_{-a}\left( x\right) ,\phi _{-b}\left( y\right) \right\} =-\delta
_{ab}\delta ^{3}\left( x-y\right) .&  \notag
\end{eqnarray}%
Next, we give the non null PB's between the primary constraints
\begin{eqnarray}
\left\{ \Gamma _{+a}\left( x\right) ,\bar{\Gamma}_{+b}\left( y\right)
\right\}  &=&-i\left( \gamma ^{+}\right) _{ab}\delta ^{3}\left( x-y\right)
~,~\ \ \ \ \   \notag \\
\left\{ \bar{\Gamma}_{+a}\left( x\right) ,\Gamma _{+b}\left( y\right)
\right\}  &=&-i\left( \gamma ^{+}\right) _{ba}\delta ^{3}\left( x-y\right) ~,
\label{d-72a} \\
\left\{ C^{k}\left( x\right) ,C^{j}\left( y\right) \right\}  &=&-2\delta
_{j}^{k}\partial _{-}^{x}\delta ^{3}\left( x-y\right) \ .  \notag
\end{eqnarray}

\subsection{The fermionic sector}

In order to the primary constraints be preserved in the time, we compute the
consistence condition of the fermionic constraints (\ref{d68-2a}). For $%
\Gamma _{+}$ we obtain
\begin{equation}
\dot{\Gamma}_{+}=\left( i\partial \!\!\!\slash-g{A\!\!\!\slash}-m\right)
\psi _{-}-gA_{+}\gamma ^{+}\psi _{+}-i\gamma ^{+}v_{1}\approx 0\ ,
\label{d72-2}
\end{equation}%
from this equation we can get two relations by using the null-plane $\gamma $%
-algebra. First we do $\displaystyle\frac{i}{2}\gamma ^{-}\dot{\Gamma}_{+}$
then we get one component of the multiplier $v_{1}$%
\begin{equation}
\Delta ^{+}v_{1}=-\frac{i}{2}\gamma ^{-}\left( i\partial \!\!\!\slash-g{%
A\!\!\!\slash}-m\right) \psi _{-}+igA_{+}\psi _{+}.  \label{qed4-15}
\end{equation}%
The second relation is obtained by the projection $\Delta ^{+}\dot{\Gamma}%
_{+}$ getting a secondary constraint
\begin{equation}
\Delta ^{+}\dot{\Gamma}_{+}=\left( i\partial \!\!\!\slash-g{A\!\!\!\slash}%
-m\right) \Delta ^{+}\psi _{-}\approx 0\ ,
\end{equation}%
due to $\left( i\partial \!\!\!\slash-g{A\!\!\!\slash}-m\right) $ is an
invertible operator we can set the secondary constraint as being
\begin{equation}
\Phi =\Delta ^{+}\psi _{-}\approx 0\ .  \label{qed4-17}
\end{equation}

For $\bar{\Gamma}_{+}$, we get
\begin{equation}
\dot{\bar{\Gamma}}_{+}=\bar{\psi}_{-}\left( i\overleftarrow{\partial \!\!\!%
\slash}+g{A\!\!\!\slash}+m\right) +gA_{+}\bar{\psi}_{+}\gamma ^{+}-i\bar{v}%
_{1}\gamma ^{+}\approx 0\ .  \label{d72-1}
\end{equation}%
From $\displaystyle\frac{i}{2}\dot{\bar{\Gamma}}_{+}\gamma ^{-}$ we get one
component of the multiplier $\bar{v}_{1}$%
\begin{equation}
\bar{v}_{1}\Delta ^{-}=-\frac{i}{2}\bar{\psi}_{-}\left( i\overleftarrow{%
\partial \!\!\!\slash}+g{A\!\!\!\slash}+m\right) \gamma ^{-}-igA_{+}\bar{\psi%
}_{+}~~,  \label{qed4-20}
\end{equation}%
and from $\dot{\bar{\Gamma}}_{+}\Delta ^{-}$ we get another secondary
constraint
\begin{equation}
\dot{\bar{\Gamma}}_{+}\Delta ^{-}=\bar{\psi}_{-}\Delta ^{-}\left( i%
\overleftarrow{\partial \!\!\!\slash}+g{A\!\!\!\slash}+m\right) \approx 0~,
\end{equation}%
similarly, $\left( i\overleftarrow{\partial \!\!\!\slash}+g{A\!\!\!\slash}%
+m\right) $ is an invertible operator, thus, we set this secondary
constraint to be
\begin{equation}
\bar{\Phi}=\bar{\psi}_{-}\Delta ^{-}\approx 0.  \label{qed4-22}
\end{equation}

The consistence condition of the constraints $\Gamma _{-}$ and $\bar{\Gamma}%
_{-}$ we get more two secondary constraints
\begin{equation}
\Omega _{-}=\dot{\Gamma}_{-}=\gamma ^{-}\left( i\partial _{-}-gA_{-}\right)
\psi _{-}+\left( i\partial \!\!\!\slash\ -g{A\!\!\!\slash}-m\right) \psi
_{+}\approx 0  \label{d73-2a}
\end{equation}%
and
\begin{equation}
\bar{\Omega}_{-}=\dot{\bar{\Gamma}}_{-}=\left( i\partial _{-}+gA_{-}\right)
\bar{\psi}_{-}\gamma ^{-}+\bar{\psi}_{+}\left( i\overleftarrow{\partial
\!\!\!\slash}+g{A\!\!\!\slash}+m\right) \approx 0\   \label{d73-1a}
\end{equation}

The consistence condition of the secondary constraints $\Phi $, $\bar{\Phi}$%
, $\Omega _{-}$ and $\bar{\Omega}_{-}$ gives relations for some components
of the fermionic multipliers, as we will show. Thus, the conservation in
time of $\Phi $ and $\bar{\Phi}$ fixes one projection for each multiplier $%
v_{2}$ and $\bar{v}_{2}$%
\begin{equation}
\dot{\Phi}=-\Delta ^{+}v_{2}\approx 0~\ \ \ \ \ ,~\ \ \ \ \ \dot{\bar{\Phi}}%
=-\bar{v}_{2}\Delta ^{-}\approx 0.  \label{qed4-26}
\end{equation}%
And the conservation in time of $\Omega _{-}$ and $\bar{\Omega}_{-}$ gives
two equations relating the multipliers $v_{2}$, $v_{1}$ and $\bar{v}_{2}$, $%
\bar{v}_{1}$, respectively,
\begin{equation}
\dot{\Omega}_{-}=-\gamma ^{-}\left( i\partial _{-}-gA_{-}\right)
v_{2}-\left( i\partial \!\!\!\slash\ -g{A\!\!\!\slash}-m\right) v_{1}\approx
0,  \label{eq-74-1}
\end{equation}%
and,
\begin{equation}
\dot{\bar{\Omega}}_{-}=-\left( i\partial _{-}+gA_{-}\right) \bar{v}%
_{2}\gamma ^{-}-\bar{v}_{1}\left( i\overleftarrow{\partial \!\!\!\slash}+g{%
A\!\!\!\slash}+m\right) \approx 0.  \label{eq-74-2}
\end{equation}

At once, we will show that the set of equation (\ref{qed4-15}), (\ref%
{qed4-20}), (\ref{qed4-26}), (\ref{eq-74-1}) and (\ref{eq-74-2}) allows to
define completely all the fermionic Lagrange multipliers. \ Thus, doing $%
\Delta ^{-}\dot{\Omega}_{-}$ we get the other component for $v_{1}$
\begin{equation}
\Delta ^{-}v_{1}=0~,
\end{equation}%
then using that $\Delta ^{+}v_{1}+\Delta ^{-}v_{1}=v_{1}$, we determine the
multiplier $v_{1}$
\begin{equation}
v_{1}=-\frac{i}{2}\gamma ^{-}\left( i\partial \!\!\!\slash-g{A\!\!\!\slash}%
-m\right) \psi _{-}+igA_{+}\psi _{+}~.  \label{qed4-30}
\end{equation}%
Also in (\ref{eq-74-1}) doing $\displaystyle\frac{1}{2}\gamma ^{+}\dot{\Omega%
}_{-}$ we determinate the component $\Delta ^{-}v_{2}$ by the equation%
\begin{equation}
\left( i\partial _{-}-gA_{-}\right) \left( \Delta ^{-}v_{2}\right) =\frac{i}{%
2}\left[ -\left( i\partial \!\!\!\slash-g{A\!\!\!\slash}\right) ^{2}+m^{2}%
\right] \psi _{-}-\frac{i}{2}gA_{+}\gamma ^{+}\left( i\partial \!\!\!\slash\
-g{A\!\!\!\slash}-m\right) \psi _{+}
\end{equation}%
the joined to component $\Delta ^{+}v_{2}$ in (\ref{qed4-26}) gives
\begin{equation}
\left( i\partial _{-}-gA_{-}\right) v_{2}=\frac{i}{2}\left[ -\left(
i\partial \!\!\!\slash-g{A\!\!\!\slash}\right) ^{2}+m^{2}\right] \psi _{-}-%
\frac{i}{2}gA_{+}\gamma ^{+}\left( i\partial \!\!\!\slash\ -g{A\!\!\!\slash}%
-m\right) \psi _{+}.  \label{4d-33}
\end{equation}

A similar procedure allows determinate the multipliers $\bar{v}_{1}$ and $%
\bar{v}_{2}$%
\begin{equation}
\bar{v}_{1}=-\frac{i}{2}\bar{\psi}_{-}\left( i\overleftarrow{\partial \!\!\!%
\slash}+g{A\!\!\!\slash}+m\right) \gamma ^{-}-igA_{+}\bar{\psi}_{+}~,
\label{4d-34}
\end{equation}%
and
\begin{equation}
\left( i\partial _{-}+gA_{-}\right) \bar{v}_{2}=\frac{i}{2}\bar{\psi}_{-}%
\left[ -\left( i\overleftarrow{\partial \!\!\!\slash}+g{A\!\!\!\slash}%
\right) ^{2}+m^{2}\right] +\frac{i}{2}gA_{+}\bar{\psi}_{+}\left( i%
\overleftarrow{\partial \!\!\!\slash}+g{A\!\!\!\slash}+m\right) \gamma ^{+}~,
\end{equation}%
respectively. Therefore, all the fermionic Lagrange multipliers were
determined, then, the set of primary and secondary fermionic constraints are
second class according to Dirac's procedure \cite{[9]}. The use of the
projection of the fermionic fields permitted to observe clearly the
existence of fermionic secondary constraints which show that the field is
fully describe by only two of their four components.

\subsection{The electromagnetic sector}

The consistent condition of $C^{k}$ gives
\begin{equation}
\dot{C}^{k}=\left( \delta _{2}^{k}\partial _{1}-\delta _{1}^{k}\partial
_{2}\right) F_{12}-g\left( \bar{\psi}_{+}\gamma ^{k}\psi _{-}+g\bar{\psi}%
_{-}\gamma ^{k}\psi _{+}\right) +\partial _{k}\pi ^{-}-2\partial
_{-}u_{k}\approx 0,  \label{d72}
\end{equation}%
it is a differential equation for the $u_{k}$ Lagrange multipliers.

The consistence condition of $\pi ^{+}$ produces the following secondary
constraint
\begin{equation}
G\equiv \dot{\pi}^{+}=\partial _{-}\pi ^{-}+\partial _{k}\pi ^{k}-g\bar{\psi}%
_{+}\gamma ^{+}\psi _{+}\approx 0,  \label{d73}
\end{equation}%
which is the Gauss's law, and its consistence condition shows that
\begin{equation}
\dot{G}=ig\left( \dot{\bar{\Gamma}}_{+}\psi _{+}+\bar{\psi}_{+}\dot{\Gamma}%
_{+}+\bar{\psi}_{-}\dot{\Gamma}_{-}+\dot{\bar{\Gamma}}_{-}\psi _{-}\right)
\approx 0\ ,  \label{d74}
\end{equation}%
what is automatically conserved in time. Then, no more constraints in the
theory are generated and the multiplier $u$ relative to $\pi ^{+}$ remains
undetermined.

\subsection{Constraint classification and gauge fixing conditions}

The full set of primary and secondary constraints is given by the equations (%
\ref{d68-1}), (\ref{d68-2a}), (\ref{qed4-17}), (\ref{qed4-22}), (\ref{d73-2a}%
), (\ref{d73-1a}) and (\ref{d73})
\begin{equation}
C\ \ ,~\ C^{k}\ \ ,~\ \Gamma _{+}\ \ ,~\ \Gamma _{-}\ \ ,~\ \bar{\Gamma}%
_{+}\ \ ,~\ \bar{\Gamma}_{-}\ \ ,~\ \Omega _{-}\ \ ,~\ \Phi \ \ ,~\ \bar{%
\Omega}_{-}\ \ ,~\ \bar{\Phi}\ \ ,~\ G~.  \label{d74a}
\end{equation}

The $C=\pi ^{+}$ has a vanishing PB with each one of the constraints and
therefore it is a first class constraint. Apparently, the remaining subset
of constraints is second class, but they form a singular constraint matrix
with an zero mode whose respective eigenvector gives a linear combination
what is one more first class constraint (see Appendix \textbf{C}).
Alternately, we must observe that as the fermionic case, the electromagnetic
sector must maintain its free constraint structure due that the interaction
term is not allowed to change the first class structure of the free theory
into second class ones, because the DB would not be defined any longer in
the limit of zero coupling constant. Thus, such combination, which is
independent of $\pi ^{+}$ and it is a first class constraint, is
\begin{equation}
\Sigma \equiv G-ig\left[ \bar{\psi}_{+}\Gamma _{+}+\bar{\Gamma}_{+}\psi _{+}+%
\bar{\psi}_{-}\Gamma _{-}+\bar{\Gamma}_{-}\psi _{-}\right] .  \label{d75}
\end{equation}

Thus, we have the following set of second class constraints
\begin{equation}
\Gamma _{+}\ \ ,\ ~\Gamma _{-}\ \ ,\ ~\Omega _{-}\ \ ,\ ~\Phi \ \ ,\ ~\bar{%
\Gamma}_{-}\ \ ,\ ~\bar{\Gamma}_{+}\ \ ,\ ~\bar{\Omega}_{-}\ \ ,\ ~\bar{\Phi}%
\ \ ,\ ~C^{k}  \label{d76}
\end{equation}%
and the set of first class constraints%
\begin{equation}
C\ \ ,\ ~\Sigma \;.  \label{d.77}
\end{equation}%
This is the maximal number of first class constraints and the consistence
condition on the second class constraints led to expressions for their
respective Lagrange multipliers.

Now, the next step is to impose gauge conditions one for every first class
constraint, such that the set of gauge fixing conditions and first class
constraints turn on a second class set. The choosing of the appropriate set
of gauge conditions is a careful procedure, because they should be
compatible with the Euler-Lagrange equations of motion. Thus, we choose a
set of gauge conditions known as the null-plane gauge and it is defined by
the following relations
\begin{equation}
B=A_{-}\approx 0\ \ \ ,\ \ \ K\equiv \pi ^{-}+\partial _{-}A_{+}\approx 0\
\label{d.78}
\end{equation}
then, the set of first class constraints and gauge fixing conditions now
\textbf{is} a second class set.

It in worthwhile to note that when the photon field is couple with the
fermion field, we would consider $A_{+}$ as a possible gauge condition but
in this case it is not possible to find a second gauge condition to be
compatible with the equations of motion. It is similar what happen with the
radiation gauge in the instant form formalism, \emph{i. e.}, $x^{0}=cte.$
plane \cite{[9]}.

\subsection{The Dirac's brackets}

The explicit evaluation of the inverse of the full matrix of second class
constraint involves an arbitrary function, which can be determined
considering appropriated boundary conditions on the fields \cite{[5]}, thus,
the inverse of the operator $\partial _{-}$ is defined as in (\ref{IC.5}).

After a laborious work, we obtain the graded Lie algebra for the dynamical
variables of the spinor electrodynamics
\begin{equation}
\left\{ A_{k}\left( x\right) ,A_{j}\left( y\right) \right\} _{D}=-\frac{1}{4}%
\delta _{j}^{k}\epsilon \left( x^{-}-y^{-}\right) \delta ^{2}\left( x^{\bot
}-y^{\bot }\right) ,  \label{DB1}
\end{equation}%
\begin{eqnarray}
\left\{ \psi _{a}\left( x\right) ,\bar{\psi}_{b}\left( y\right) \right\}
_{D} &=&-\frac{i}{2}~\left( \gamma ^{-}\right) _{ab}\delta ^{3}\left(
x-y\right) -\frac{1}{4}~\epsilon \left( x^{-}-y^{-}\right) ~  \notag \\
&&\left( iD\!\!\!\!\slash_{\bot }^{\;x}{+m}\right) _{ab}\delta ^{2}\left(
x^{\bot }-y^{\bot }\right) +\frac{i}{8}~\left\vert x^{-}-y^{-}\right\vert
\label{DB2} \\
&&\left\{ \gamma ^{+}\left[ \left( D\!\!\!\!\slash_{\bot }^{~x}\right) ^{2}{%
+m}^{2}\right] \right\} _{ab}\delta ^{2}\left( x^{\bot }-y^{\bot }\right) ,
\notag
\end{eqnarray}%
\begin{eqnarray}
\left\{ A_{+}\left( x\right) ,\psi \left( y\right) \right\} _{D} &=&-\frac{i%
}{2}g\left\vert x^{-}-y^{-}\right\vert \delta ^{2}\left( x^{\bot }-y^{\bot
}\right) \psi \left( y\right)   \label{DB3} \\[0.3cm]
&&+\frac{i}{4}g\delta ^{2}\left( x^{\bot }-y^{\bot }\right) \int
\!\!dz^{-}\epsilon \left( x^{-}-z^{-}\right) \epsilon \left(
z^{-}-y^{-}\right) \Delta ^{-}\psi \left( x^{\bot },z^{-}\right)   \notag \\%
[0.3cm]
&&+\frac{i}{16}g\delta ^{2}\left( x^{\bot }-y^{\bot }\right) \int
\!\!dz^{-}~\left\vert x^{-}-z^{-}\right\vert \epsilon \left(
z^{-}-y^{-}\right) ~\left[ \gamma ^{+}\gamma ^{k}\partial _{k}^{x}\psi
\left( x^{\bot },z^{-}\right) \right]   \notag \\[0.3cm]
&&+\frac{i}{16}g\left[ \partial _{k}^{x}\delta ^{2}\left( x^{\bot }-y^{\bot
}\right) \right] \int \!\!dz^{-}~\left\vert x^{-}-z^{-}\right\vert \epsilon
\left( z^{-}-y^{-}\right) \left[ \gamma ^{+}\gamma ^{k}\psi \left( x^{\bot
},z^{-}\right) \right] ,  \notag
\end{eqnarray}%
\begin{eqnarray}
\left\{ A_{+}\left( x\right) ,\bar{\psi}\left( y\right) \right\} _{D}
&=&\strut \frac{i}{2}g\left\vert x^{-}-y^{-}\right\vert \delta ^{2}\left(
x^{\bot }-y^{\bot }\right) ~\bar{\psi}\left( y\right)   \label{DB4} \\[0.3cm]
&&-\frac{i}{4}g\delta ^{2}\left( x^{\bot }-y^{\bot }\right) \int
\!\!dz^{-}\epsilon \left( x^{-}-z^{-}\right) \epsilon \left(
z^{-}-y^{-}\right) \bar{\psi}\left( x^{\bot },z^{-}\right) \Delta ^{+}
\notag \\[0.3cm]
&&-\frac{i}{16}g\left[ \partial _{j}^{x}\delta ^{2}\left( x^{\bot }-y^{\bot
}\right) \right] \int \!\!dz^{-}~\left\vert x^{-}-z^{-}\right\vert \epsilon
\left( z^{-}-y^{-}\right) \left[ \bar{\psi}\left( x^{\bot },z^{-}\right)
\gamma ^{j}\gamma ^{+}\right]   \notag \\[0.3cm]
&&-\frac{i}{16}g\delta ^{2}\left( x^{\bot }-y^{\bot }\right) \int
\!\!dz^{-}\ \left\vert x^{-}-z^{-}\right\vert \epsilon \left(
z^{-}-y^{-}\right) \left[ \partial _{k}^{x}\bar{\psi}\left( x^{\bot
},z^{-}\right) \gamma ^{k}\gamma ^{+}\right] .  \notag
\end{eqnarray}

Our two first expressions, equations (\ref{DB1}) and (\ref{DB2}), are in
accord with the result obtained by Rohrlich \cite{[5]} and Kogut and Soper
\cite{[14]} when the correspondence principle is applied. In \cite{[5]}
these relations are determined from the (anti)-commutator between the field
operators for unequal times while in \cite{[14]} the authors postulated the
(anti)-commutators for the annihilation and creation operators for the
quanta fields, and transforming these relations back to coordinate space,
they obtained the equal-time (anti)-commutators.

The relations (\ref{DB3}) and (\ref{DB4}) associated with the field $A_{+}$
were derived following a careful application of the Dirac's procedure to the
null-plane. Using the correspondence principle these relations are
equivalents with the expressions derived by Kogut and Soper \cite{[14]}.

\smallskip

\section{Remarks and conclusions}

We have performed the constraint analysis of the scalar and spinor
electrodynamics on the null-plane and several characteristics or features
are in contrast with the customary space-like hyper-surface formulation.

We have shown that the $SQED_{4}$ has a first class constraint, the Gauss's
law, which result of a linear combination of electromagnetic and scalar
constraints which is given by the zero mode eigenvector of the constraint
matrix. This fact is a consequence of the constraints associated with the
scalar sector. On the other hand, in the instant form analysis the second
first class constraint does not have contributions coming from scalar
constraints because in this formalism the scalar sector is free of
constraints.

After select the null-plane gauge conditions to transform the first class
constraints in second class one, we need to impose appropriated boundary
conditions on the fields to fix a hidden subset of first class constraints
which allows to get an unique inverse of the second class constraints
matrix. Then, we obtain the DB's of the theory and can quantize it via the
correspondence principle. The commutation relations among fields (\ref{A})
obtained by us are consistent with those results reported in the literature
\cite{[12]}. The relations (\ref{C}) involving the field operator ${A}_+$
were not obtained in \cite{[12]} but it was affirmed that they could be
derived solving a quantum constraint. However, these commutators were
calculated by us by quantizing the DB's derived at classical level following
a careful analysis of the constraints structure of the $SQED_{4}$.

In $QED_{4}$ case, the careful analysis the constraints structure of the
fermionic sector shows that it has only second class constraints. However,
there exists a hidden subset of first class constraints \cite{[7]} which
generate improper gauge transformations \cite{[8]}. Such first class subset
are associated with the impossibility of define an unique inverse for the
operator $\partial _{-}$ related to the insufficiency of boundary conditions
to solve the Cauchy data problem. The uniqueness of the inverse is
guaranteed by imposing appropriated boundary conditions on the fields.

The first class set of the $QED_{4}$ are fixed choosing the null-plane gauge
conditions and following the Dirac's procedure we obtain graded algebra (\ref%
{DB1})-(\ref{DB4}) for the canonical variables. Via the correspondence
principle (\ref{DB1}) and (\ref{DB2}) reproduce the canonical
(anti)-commutation relations for the quantum fields derived in \cite%
{[5],[14]}. Also the relations (\ref{DB3}) and (\ref{DB4}) associated with
the field $A_{+}$, derived following a careful application of the Dirac's
procedure, reproduce the expressions derived by Kogut and Soper \cite{[14]}
at quantum level.

Recently \cite{hadrons} the null-plane Hamiltonian structure of the
(1+1)-dimensional electrodynamics, the Schwinger model, has been studied by
following the Dirac's procedure \cite{[2]} and dealing carefully the hidden
first class constraints \cite{[7],[8]}. The study shows that the fermionic
sector has only second class constraints such as happened in the instant
formalism. And as we can be shown the fermionic sector of the $QED_{4}$ in
the null-plane also presents only second class constraints structure. It can
be conclude that fermionic fields satisfying a first order Dirac's equation
in the null-plane or instant form formalisms have only second class
constraints structure.

In advanced, we are studying the constraints structure analysis of the pure
Yang-Mills fields and also the Hamiltonian structure of the theory resulting
of the interaction between the Yang-Mils fields with complex scalar fields.
Reports on this research will be communicated elsewhere.

\subsection*{Acknowledgements}

BMP thanks CNPq for partial support. GERZ thanks CNPq (grant 142695/2005-0)
for full support.

\appendix

\section{Notation}

The null plane time $x^{+}$ and longitudinal coordinate $x^{-}$ are defined,
respectively, as
\begin{equation}
x^{+}\equiv\frac{x^{0}+x^{3}}{\sqrt{2}}\,\ \ \ \ \ \ \ \ x^{-}\equiv \frac{%
x^{0}-x^{3}}{\sqrt{2}},  \label{A2.1}
\end{equation}
with the transverse coordinates $x^{\bot }\equiv(x^{1},x^{2})$ kept
unchanged.

Hence, in the space of four-vectors $x=(x^{+},x^{1},x^{2},x^{-})$, the
metric is
\begin{equation}
g={\left(
\begin{array}{cccc}
0 & 0 & 0 & 1 \\
0 & -1 & 0 & 0 \\
0 & 0 & -1 & 0 \\
1 & 0 & 0 & 0%
\end{array}
\right) }\ \ .  \label{A2.2}
\end{equation}

Explicitly,
\begin{equation}
x^{+}=x_{-}\ ,\ \ \ x^{-}=x_{+}\ \ ,\ \ \ x\cdot
y=x^{+}y^{-}+x^{-}y^{+}-x^{\bot }\cdot y^{\bot }\;\ ,  \label{A2.4}
\end{equation}%
where the derivatives with respect to $x^{+}$ e $x^{-}$ are defined as
\begin{equation}
\partial _{+}\equiv \frac{\partial }{\partial x^{+}}\ ,\ \ \ \ \ \ \partial
_{-}\equiv \frac{\partial }{\partial x^{-}}  \label{A2.6}
\end{equation}%
with $\partial ^{+}=\partial _{-}$. Here, we have used the following
relations
\begin{equation*}
\delta ^{4}(x-y)=\delta (x^{+}-y^{+})\delta ^{2}(x^{\perp }-y^{\perp
})\delta (x^{-}-y^{-}).
\end{equation*}%
\begin{equation}
\frac{1}{2}\frac{d}{dx^{-}}\epsilon (x^{-}-y^{-})=\delta (x^{-}-y^{-})\ \ \
\ ,\ \ \ \ \frac{1}{2}\int\!\! dy^{-}~\epsilon (x^{-}-y^{-})\epsilon
(y^{-}-z^{-})=|x^{-}-y^{-}|  \label{A2.9}
\end{equation}

The same orthogonal transformation is applied to Dirac matrices that still
obey
\begin{equation}
\{\gamma^{\mu},\gamma^{\nu}\}=2g^{\mu\nu}  \label{A2.11}
\end{equation}%
this makes $\gamma^{+}$ and $\gamma^{-}$ singular matrices.

Since
\begin{equation}
(\gamma ^{+})^{\dag }=\gamma ^{-}\ ,\ \ \ \ \ \ (\gamma ^{-})^{\dag }=\gamma
^{+}\,\ \ \ \ \ \ (\gamma ^{k})^{\dag }=-\gamma ^{k}\qquad k=1,2.
\label{A2.13}
\end{equation}%
we define the Hermitian matrices
\begin{equation}
\Delta ^{\pm }=\frac{1}{2}\gamma ^{\mp }\gamma ^{\pm }\;,  \label{A2.14}
\end{equation}%
which are projection operators,
\begin{equation}
(\Delta ^{\pm })^{2}=\Delta ^{\pm }\ ,\ \ \ \ \ \Delta ^{\pm }\Delta ^{\mp
}=0\ ,\ \ \ \ \ \ \Delta ^{+}+\Delta ^{-}=1\ \ .  \label{A2.15}
\end{equation}%
Their action on Dirac spinors yields
\begin{equation}
\psi _{\pm }=\Delta ^{\pm }\psi \ ,\ \ \ \ \ \ \ \ \bar{\psi}_{\pm }=\bar{%
\psi}\Delta ^{\mp },  \label{A2.16}
\end{equation}

\section{The second first class constraint for SQED and QED}

Now, we compute the second first class constraint for $SQED_{4}$. The set of
the remaining constraints is $\Phi ^{a}=\left\{ \Gamma ,~\Gamma ^{\ast
},~G,~\chi ^{k}\right\} $, its constraint matrix $C_{ab}\left( x,y\right)
=\left\{ \Phi ^{a}\left( x\right) ,\Phi ^{a}\left( y\right) \right\} =\Delta
\left( x\right) \delta ^{3}\left( x-y\right) $, where the $\Delta \left(
x\right) $ matrix is
\begin{equation}
\Delta \left( x\right) =\left(
\begin{array}{cccc}
0 & -2\left( D_{-}^{x}\right) ^{\ast } & 2ig\left[ \left( D_{-}^{x}\right)
^{\ast }\phi \left( x\right) \right] +2ig\phi \left( x\right) \partial
_{-}^{x} & 0 \\[0.3cm]
-2D_{-}^{x} & 0 & -2ig\left[ D_{-}^{x}\phi ^{\ast }\left( x\right) \right]
-2ig\phi ^{\ast }\left( x\right) \partial _{-}^{x} & 0 \\[0.25cm]
2ig\phi \left( x\right) D_{-}^{x} & -2ig\phi ^{\ast }\left( x\right) \left(
D_{-}^{x}\right) ^{\ast } & F\left( x\right) & 0 \\[0.25cm]
0 & 0 & 0 & -2\delta _{l}^{k}\partial _{-}^{x}%
\end{array}%
\right)
\end{equation}%
where the operator $F$ is
\begin{equation}
F\equiv -2g^{2}\partial _{-}\left( \phi \phi ^{\ast }\right) -4g^{2}\phi
\phi ^{\ast }\partial _{-}.
\end{equation}

The matrix $C\left( x,y\right) $ has determinant zero. It is due to this
matrix has a zero eigenvalue and its respective eigenvector gives a linear
combination of the constraints which is a first class constraint. The
eigenvector is calculated using the following equation%
\begin{equation}
\int d^{3}y~C_{ab}\left( x,y\right) U_{b}\left( y\right) =0,
\end{equation}%
thus, the eigenvector is $U=\left( -ig\phi ^{\ast },~ig\phi ,~1,~0,~0\right)
^{T}$ and the linear combination given the second first class is
\begin{equation}
\Sigma =\Phi ^{a}U_{a}=G-ig\left( \phi ^{\ast }\Gamma -\phi \Gamma ^{\ast
}\right) .
\end{equation}

For the $QED_{4}$, the remaining set of constraints is
\begin{equation}
\ ~\ G\ \ ,~\ \Gamma _{+}\ \ ,~\ \Gamma _{-}\ \ ,~\ \bar{\Gamma}_{+}\ \ ,~\
\bar{\Gamma}_{-}\ \ ,~\ \Omega _{-}\ \ ,~\ \Phi \ \ ,~\ \bar{\Omega}_{-}\ \
,~\ \bar{\Phi}\ \ ,~\ C^{k}.
\end{equation}%
Its constraint matrix has a sole zero mode and the respective eigenvector is
\begin{equation}
U=(1,~ig\bar{\psi}_{+},~ig\bar{\psi}_{-},~-ig\psi _{+},-ig\psi
_{-},~0,~0,~0,~0,~0,~0)^{T},
\end{equation}%
it gives the second first class
\begin{equation}
\Sigma =G-ig\left[ \bar{\psi}_{+}\Gamma _{+}+\bar{\Gamma}_{+}\psi _{+}+\bar{%
\psi}_{-}\Gamma _{-}+\bar{\Gamma}_{-}\psi _{-}\right] .
\end{equation}

\section{Grassmann Algebras}

A Grassmann algebra contains bosonic (self-commuting) and fermionic
(self-anticommuting) variables {\ \cite{[15]}}:
\begin{equation}
FB=(-1)^{n_{A}n_{B}}BF\ \ ,  \label{B1}
\end{equation}%
where $n=0$ for a bosonic, and $n=1$ for a fermionic variable. Note that the
product of two fermionic variables is bosonic, and the product of a
fermionic and a bosonic variables is fermionic.

The left derivative of a $\psi _{\alpha }$ fermionic variable is defined as
\begin{equation}
\frac{\partial }{\partial \psi _{\alpha }}\left\{ \frac{{}}{{}}\psi _{\alpha
_{1}}\psi _{\alpha _{2}}\ \cdot \cdot \cdot \psi _{\alpha _{n}}\right\}
=-\delta _{\alpha \alpha _{1}}\psi _{\alpha _{2}}\ \cdot \cdot \cdot \psi
_{\alpha _{n}}+\delta _{\alpha \alpha _{2}}\psi _{\alpha _{1}}\psi _{\alpha
_{3}}\ \cdot \cdot \cdot \psi _{\alpha _{n}}+\ \cdot \cdot \cdot \
+(-1)^{n}\delta _{\alpha \alpha _{n}}\psi _{\alpha _{1}}\psi _{\alpha _{2}}\
\cdot \cdot \cdot \psi _{\alpha _{n-1}}\ \ .  \label{B2}
\end{equation}

The Poisson Brackets can be defined similar to ordinary mechanics \cite{[16]}%
. The phase space is spanned by $q_{i}$, $p^{i}$ which are bosons and $\psi
_{\alpha }$ and $\pi ^{\alpha }$, fermions. Denote by B(F) a bosonic
(Fermionic) element of the Grassmann algebra, then
\begin{eqnarray}
\{B_{1},B_{2}\}=-\{B_{2},B_{1}\} &=&\left\{ \frac{\partial B_{1}}{\partial
q_{i}}\frac{\partial B_{2}}{\partial p^{i}}-\frac{\partial B_{2}}{\partial
q_{i}}\frac{\partial B_{1}}{\partial p^{i}}\right\} +\left\{ \frac{\partial
B_{1}}{\partial \phi _{\alpha }}\frac{\partial B_{2}}{\partial \pi ^{\alpha }%
}-\frac{\partial B_{2}}{\partial \phi _{\alpha }}\frac{\partial B_{1}}{%
\partial \pi ^{\alpha }}\right\}  \notag  \label{B4} \\[0.2cm]
\{F,B\}=-\{B,F\} &=&\left\{ \frac{\partial F}{\partial q_{i}}\frac{\partial B%
}{\partial p^{i}}-\frac{\partial B}{\partial q_{i}}\frac{\partial F}{%
\partial p^{i}}\right\} -\left\{ \frac{\partial F}{\partial \phi _{\alpha }}%
\frac{\partial B}{\partial \pi ^{\alpha }}+\frac{\partial B}{\partial \phi
_{\alpha }}\frac{\partial F}{\partial \pi ^{\alpha }}\right\} \\[0.2cm]
\{F_{1},F_{2}\}=\{F_{2},F_{1}\} &=&\left\{ \frac{\partial F_{1}}{\partial
q_{i}}\frac{\partial F_{2}}{\partial p^{i}}+\frac{\partial F_{2}}{\partial
q_{i}}\frac{\partial F_{1}}{\partial p^{i}}\right\} -\left\{ \frac{\partial
F_{1}}{\partial \phi _{\alpha }}\frac{\partial F_{2}}{\partial \pi ^{\alpha }%
}+\frac{\partial F_{2}}{\partial \phi _{\alpha }}\frac{\partial F_{1}}{%
\partial \pi ^{\alpha }}\right\} \ \ .  \notag
\end{eqnarray}%
It follow from its definition that the Poisson brackets has the properties
\begin{eqnarray}
\{A,B\} &=&-(-1)^{n_{A}n_{B}}\{B,A\}  \notag  \label{B5} \\[0.2cm]
\{A,B+C\} &=&\{A,B\}+\{A,C\}  \notag \\[0.2cm]
\{A,BC\} &=&(-1)^{n_{A}n_{B}}B\{A,C\}+\{A,B\}C \\[0.2cm]
\{AB,C\} &=&(-1)^{n_{B}n_{C}}\{A,C\}B+A\{B,C\}  \notag \\[0.2cm]
(-1)^{n_{A}n_{C}}\{A,\{B,C\}\}
&+&(-1)^{n_{B}n_{A}}\{B,\{C,A\}\}+(-1)^{n_{C}n_{B}}\{C,\{A,B\}\}=0\ \ .
\notag
\end{eqnarray}




\end{document}